\documentclass[a4paper,fleqn,usenatbib]{mnras}

\usepackage{newtxtext,newtxmath}
\usepackage[T1]{fontenc}
\usepackage{ae,aecompl}
\usepackage{graphicx}   % Including figure files
\usepackage{amsmath}    % Advanced maths commands
\usepackage{amssymb}    % Extra maths symbols
\usepackage{makecell}   % Make the text in the tables centered
\usepackage{threeparttable} % Set one format of the tables
\usepackage{multirow}   % Set one format of the tables  

\title[BALs that consist of blended NALs]{Narrow absorption lines complex I: one form of broad absorption line}

\author[W.-J. Lu et al.]{
Wei-Jian Lu,\footnotemark[1]
Ying-Ru Lin\footnotemark[1]
\\
School of Information Engineering, Baise University, Baise 533000, China}

\date{Accepted XXX. Received YYY; in original form ZZZ}

\pubyear{2017}

\begin{document}
\label{firstpage}
\pagerange{\pageref{firstpage}--\pageref{lastpage}}
\maketitle

\begin{abstract}
We discover that some of the broad absorption lines (BALs) are actually a complex of narrow absorption lines (NALs). As a pilot study of this type of BAL, we show this discovery through a typical example in this paper. Utilizing the two-epoch observations of J002710.06-094435.3 (hereafter J0027-0944) from the Sloan Digital Sky Survey (SDSS), we find that each of the \ion{C}{iv} and \ion{Si}{iv} BAL troughs contains at least four NAL doublets. By resolving the \ion{Si}{iv} BAL into multiple NALs, we present the following  main results and conclusions. First, all these NALs show coordinated variations between the two-epoch SDSS observations, suggesting that they all originate in the quasar outflow, and that their variations are due to global changes in the ionization condition of the absorbing gas. Secondly, a BAL consisting of a number of NAL components indicates that this type of BAL is basically the same as the intrinsic NAL, which tends to support the inclination model rather than the evolution model. Thirdly, although both the \ion{C}{iv} and \ion{Si}{iv} BALs originate from the same clumpy substructures of the outflow, they show different profile shapes: multiple absorption troughs for the \ion{Si}{iv} BAL in a wider velocity range, while P-Cygni for the \ion{C}{iv} BAL in a narrower velocity range. This can be interpreted by the substantial differences in fine structure and oscillator strength between the \ion{Si}{iv}$\lambda\lambda$1393, 1402  and \ion{C}{iv}$\lambda\lambda$1548, 1551 doublets. Based on the above conclusions, we consider that the decomposition of a BAL into NALs can serve as a way to resolve the clumpy structure for outflows, and it can be used to learn more about characteristics of the clumpy structure and to test the outflow model, when utilizing high-resolution spectra and photoionization model.

\end{abstract}
%\\\ion{C}{iv} $\mathrm{\lambda\lambda}1548,1551$
%$\rm km~s^{-1}$
%$\rm P\hat{a}ris$ et al.2012

\footnotetext[1]{E-mail: william\_lo@qq.com (W-JL); yingru\_lin@qq.com (Y-RL)}

%%%%%%%%%%%%%%%%%%%%%%%%%%%%%%%%%%%%%%%%%%%%%%%%%%

%%%%%%%%%%%%%%%%% BODY OF PAPER %%%%%%%%%%%%%%%%%%

\section{Introduction}
Quasar spectra show many rest-frame ultraviolet (UV) absorption lines. According to their distances from the background quasar, absorption lines can be divided into two classes: (1) intrinsic absorption lines that are caused by the gases inside the quasar; (2) intervening absorption lines that are caused by the gas in/around the quasar host galaxy, galaxy cluster around the quasar, or the foreground media having no physical relations with the quasar. The intrinsic absorption lines are usually classified into three categories according to their absorption widths: broad absorption lines (BALs: absorption widths of at least 2000 $\rm km~s^{-1}$; Weymann et al. 1991) detected in about 41 per cent of quasars (Allen et al. 2011); mini-BALs (absorption widths from 500 to 2000 $\rm km~s^{-1}$; Hamann \& Sabra 2004); narrow absorption lines (NALs: absorption widths of less than 500 $\rm km~s^{-1}$) detected in $\thicksim20-50$ per cent of quasars (e.g. Misawa et al. 2007).

At present, the relationship between BAL, mini-BAL and intrinsic NAL is still unclear. In an evolution model picture (e.g. Farrah et al. 2007), different types of intrinsic absorption lines may represent different evolution stages of the quasar outflows. For example, the BAL might represent a powerful phase of the outflow, while the intrinsic NAL and mini-BAL are the beginning or the ending of the BAL outflow (e.g. Hamann et al. 2008). In an inclination model picture (e.g. Murray et al. 1995; Elvis 2000; Proga, Stone \& Kallman 2000), different types of intrinsic absorption lines may be due to viewing angle effects. For instance, the BAL might represent the main body of the outflow, while the NAL and mini-BAL are the clumpy structures on the edge of the outflow at high latitudes (e.g. Ganguly et al. 2001a). 
%Figure 1----------------------------------------------
\begin{figure*}
    \includegraphics[width=2\columnwidth]{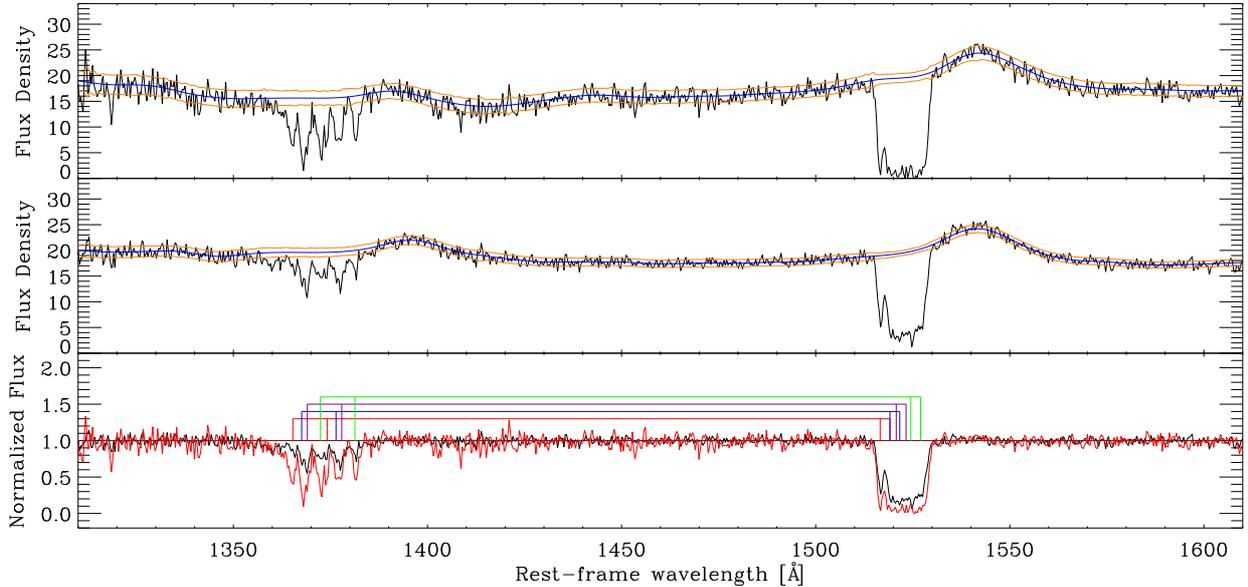}
  \caption{Spectra of quasar J0027--0944. The top and middle panels show the spectra observed by SDSS on MJD 52145 and 54825, respectively. The flux density is in units of $\rm 10^{-17} erg~ cm^{-2}~s^{-1}~$\AA$^{-1}$. The blue lines are the corresponding pseudo-continua. The orange lines represent the pseudo-continuum fluxes taking into account the flux uncertainties. In the bottom panel, the black and red lines show the normalized spectra of J0027--0944 from observations on MJD 54825 and 52145, respectively. The red, blue, purple and green lines mark out four identified NAL systems.}
    \label{fig.1}
\end{figure*}
%----------------------------

BALs exhibit diverse shapes; for example, multiple absorption troughs (e.g. Turnshek et al. 1980), detached absorption troughs (e.g. Osmer \& Smith 1977), and classic P-Cygni (e.g. Scargle, Caroff \& Noerdlinger 1970), etc. The cause of BAL profile diversity is also under debate. Hydrodynamic simulation performed by Pereyra (2014) has suggested that the discontinuities in the ionization balance of the outflow, which is caused by X-ray shielding, may result in a profile shape of multiple absorption troughs. Assuming an X-ray shielded region, Pereyra has further found that the diversity of BAL profile can be explained by the viewing angel effect: from `face-on' to `edge-on', one will successively detect multiple absorption troughs, then detached absorption troughs, and then classic P-Cygni. In an observational view, Baskin, Laor \& Hamann (2013) have found that the velocity of \ion{C}{iv} BAL profile is controlled by the \ion{He}{ii} emission equivalent width (EW), while its profile depth is controlled by the spectral slope in the 1700--3000 \AA~range. They suggested that the \ion{He}{ii} emission EW and the spectral slope may indicate the ionizing continuum and the viewing angle, respectively.

We happened to discover that a number of BALs are actually the mixtures of NALs, during a programme to study the BAL variation with time  (Lu, Lin \& Qin 2017). Through a careful visual inspection of about 2000 BAL quasars with multi-epoch observations from SDSS-I/II/III, we confirmed that this type of BALs are not rare. This interesting discovery may offer a new perspective to study outflows, for example the relationship between different types of absorption lines (BAL, mini-BAL or NAL), and the cause of the profile diversity of BALs, etc. As a pilot study of this type of BAL, we show a typical example  with two-epoch observations of quasar SDSS J002710.06-094435.3 (hereafter J0027-0944) in this paper. As a BAL quasar, J0027-0944 has been studied in many systematic studies (Trump et al. 2006; Gibson et al. 2009a; Scaringi et al. 2009; Allen et al. 2011; He et al. 2015, 2017). J0027-0944 has been observed twice by SDSS. The first-epoch SDSS spectrum of J0027-0944 has the balnicity index (BI; defined as ${\rm BI}=\int_{-25000~\rm km~s^{-1}}^{-3000~\rm km~s^{-1}} (1-\frac{f(v)}{0.9})C{\rm d}v$\footnotemark; Weymann et al. 1991) of 433.1 and 758.9 $\rm km~s^{-1}$ for the \ion{Si}{iv} and \ion{C}{iv} BALs (taken from Allen et al. 2011), respectively. The \ion{C}{iv} BAL has a P-Cygni shape, while the \ion{Si}{iv} BAL shows multiple absorption troughs. In this paper, we present that the \ion{C}{iv} and \ion{Si}{iv} BALs in J0027-0944 each actually contains at least four NAL systems. 

The paper is structured as follows. Section 2 presents the quasar spectra and describes the spectral analysis. Section 3 contains the results and discussions. Section 4 gives the conclusion and future works.

\footnotetext{Where $f (v)$ is the continuum-normalized spectral flux as a function of a velocity $v$ (in $\rm km~s^{-1}$), relative to the quasar rest frame. The dimensionless value $C$ is set to 1 where the normalized flux starts to continuously fall at least 10 per cent below the continuum for at least 2000 $\rm km~s^{-1}$, and is switched to zero everywhere else.}
%Table1-------------------------
\begin{table*}
    \centering
    \caption{Measurements of \ion{Si}{iv} and \ion{C}{iv} BALs.}
    \label{tab.1}
    \begin{tabular}{ccccrccr} 
        \hline\noalign{\smallskip}
Species & $\rm z_{abs}$ & Velocity &\multicolumn{2}{c}{MJD: 52145} &{ } & \multicolumn{2}{c}{MJD: 54825}\\
\cline{4-5}\cline{7-8}\noalign{\smallskip}
{}&{}& &$\rm EW$ & $\rm FWHM$ & {} &$\rm EW$ & $\rm FWHM$ \\
{}&{}&$({\rm km~s}^{-1})$& (\AA) & (${\rm km~s}^{-1}$)  & {}& (\AA) & (${\rm km~s}^{-1}$)\\
\hline\noalign{\smallskip}
\ion{Si}{iv}$\lambda$1393 &\multirow{2}{*}{2.0208}&\multirow{2}{*}{$-6199$}&	$1.13\pm0.29$ &	368.89& {}&$0.30\pm0.39$	  &301.80\\
\ion{Si}{iv}$\lambda$1402	& {} & {}  & $0.66\pm0.25$	  & 249.90 & {} &$0.25\pm0.25$   &199.91\\
\ion{Si}{iv}$\lambda$1393	&\multirow{2}{*}{2.0262}	&\multirow{2}{*}{$-5660$}&$1.22\pm0.18$	  &284.54 & {}  &	$0.40\pm0.23$	 & 251.08\\
\ion{Si}{iv}$\lambda$1402	&	{}    &&$0.73\pm0.32$  &299.34& {}  &	$0.33\pm0.28$  & 249.47\\
\ion{Si}{iv}$\lambda$1393	&\multirow{2}{*}{2.0290}	&\multirow{2}{*}{$-5383$}&$0.73\pm0.24$  &	250.83& {} &$0.53\pm0.15$	   &250.81\\
\ion{Si}{iv}$\lambda$1402&	{}	   &{}	   &$0.66\pm0.37$  &282.45& {} &	$0.49\pm0.18$	 &249.20\\
\ion{Si}{iv}$\lambda$1393	&\multirow{2}{*}{2.0369}	&\multirow{2}{*}{$-4603$}&$1.25\pm0.21$  &	 333.57 & {}  &	$0.43\pm0.35$	 &333.60\\
\ion{Si}{iv}$\lambda$1402&	{}	   &{}&$0.90\pm0.35$  &	314.86& {}	&$0.32\pm0.27$	  &248.59\\
\ion{C}{iv} BAL	 &--   &$-6745\thicksim-3793^{\rm a}$&$12.28\pm1.07$ &$2952^{\rm b}$& {}	&$9.74\pm0.54$  	&$2952^{\rm b}$\\
\ion{Si}{iv} BAL &--   &$-7985\thicksim-3511^{\rm a}$&$7.97\pm1.92$&$4474^{\rm b}$& &$3.66\pm0.87$&$4474^{\rm b}$\\
 \hline
    \end{tabular}
\begin{tablenotes}
\footnotesize
\item$^{\rm a}$Velocity range of the BAL troughs with respect to emission rest frame.
\item$^{\rm b}$Total width calculated from edge-to-edge of the BAL troughs.
\end{tablenotes}
 \end{table*}
%Table1-------------------------
\section{SPECTROSCOPIC ANALYSIS}% 2 %%%%%%%%%%%%%%%%%%%%%%%%%%%%%%%%%%%%%

The SDSS uses a 2.5-m Ritchey-Chretien telescope (Gunn et al. 2006) at Apache Point Observatory, New Mexico. SDSS-I/II (the first two periods of the SDSS project) spectra have a spectral resolution of $R \thickapprox 1800-2200$ (e.g. York et al. 2000). J0027--0944 ($\rm z=2.0839$, taken from Hewett \& Wild 2010) was observed by SDSS on MJD 52145 and 54825, respectively. These two observations span about 7 yr in the observed frame ($\Delta \rm{t_{obs}} = 2680~days$), i.e., about 2.4 yr in the quasar rest frame ($\Delta \rm {t_{rest}} = 869~days$). The median signal-to-noise ratio (S/N) of the MJD 52145 and MJD 54825 spectra of J0027--0944 are 16.38 and 27.46 per pixel, respectively.

 %Figure 2----------------------------------------------
\begin{figure}
\includegraphics[width=\columnwidth]{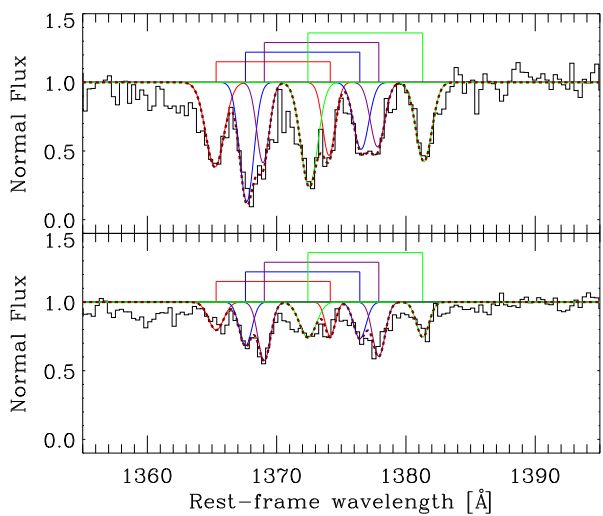}
\caption{Pseudo-continuum normalized spectra of quasar J0027--0944 with a wavelength range from 1355 to 1395 \AA~(quasar frame). The upper and lower spectra are snippets from the spectra observed on MJD 52145 and 54825, respectively. The Gaussian fittings in red, blue, purple and green represent the four identified \ion{Si}{iv} NAL systems. The red, blue, purple and green solid lines mark out four identified NAL systems. The total fit model is also marked out by a brown broken line.}
    \label{fig.2}
\end{figure}
%---------------------------- 

We fit the continua using the cubic spline functions iteratively. In order to reduce the effects of absorption troughs and remaining sky pixels, during our fitting, we masked out the pixels that beyond $3\sigma$ significance from the current fit. Our continuum fitting is shown in Fig. \ref{fig.1}. Then we measured the absorption lines in the pseudo-continuum normalized spectra. 

For the blended NALs within the \ion{Si}{iv} BAL trough, we employed four pairs of Gaussian functions to fit them (Fig. \ref{fig.2}). Limited by the low resolution of the spectra, we cannot measure or discuss the coverage fraction of NALs. Therefore, we accepted a full coverage during line fitting in this work. For each \ion{Si}{iv}  NAL doublet, we adopted the absorption redshift measured from the spectrum observed on MJD 52145 (Table \ref{tab.1}). The rest-frame EW of each NAL was measured based on the Gaussian fitting. Imitating the optimal extraction method (see Schneider et al. 1993), the error on EW was estimated using

%公式1---------------------
\begin{equation}
	    \sigma_{\rm EW}=\frac{\sqrt{\sum_{i=1}^{N} P^{2}(\lambda_i-\lambda_0)\sigma_{f_i}^{2}\Delta \lambda_{i}^2}}{(1+z_{\rm abs})\times\sum_{i=1}^{N} P^{2}(\lambda_i-\lambda_0)} ,
    \label{eq:quadratic 1}
\end{equation}
%————————————————————————
where $P(\lambda_i-\lambda_0)$ represents the Gaussian line profile centred at $\lambda_0$ and $\sigma_{f_i}$ represents the normalized flux uncertainty. $\Delta \lambda_{i}$ is a pixel scale in a unit of \AA. The sum was performed on an integer number of pixels covering $\pm3$ characteristic Gaussian widths.

For the \ion{C}{iv} BAL, we could not identify the independent NALs due to their severe blending. So we just measured the EW for the whole \ion{C}{iv} BAL trough from the normalized spectra using
 %公式2---------------------
\begin{equation}
        {\rm EW}=\sum_i (1-\frac{F_i}{F_c})\Delta \lambda_{i},
    \label{eq:quadratic 2}
\end{equation}
%————————————————————————
and the uncertainty on the EW is
%公式3---------------------
\begin{equation}
 \sigma_{\rm EW}=\sqrt{[\frac{\Delta F_c}{F_c}\sum_i (\frac{\Delta \lambda_{i}F_i}{F_c})]^2+\sum_i (\frac{\Delta \lambda_{i}\Delta F_i}{F_c})^2},
    \label{eq:quadratic 3}
\end{equation}
%————————————————————————
where $F_i$, $\Delta F_i$, $F_c$ and $\Delta F_c$ are the flux in the $i$th bin, the error on $F_i$, the underlying continuum flux, and the uncertainty in the mean continuum flux in the normalization window, respectively (Kaspi et al. 2002). In the normalized spectra, $F_c$= 1. We calculated $\Delta F_c$ using a window of 1485--1515 \AA, which is the closest window to the \ion{C}{iv} BAL trough. We also measured the EW for the whole \ion{Si}{iv} BAL. Measurements of the \ion{C}{iv} and \ion{Si}{iv} BAL are also listed in Table \ref{tab.1}.

\section{Discussion}% 3 %%%%%%%%%%%%%%%%%%%%%%%%%%%%%%%%%%%%%%%%%%%%%%                                                                                                                                                                                                                                                                                                                                                                                                          
\subsection{Origin and variability cause}% 3.1 %%%%%%%%%%%%%%%

The complex of high-redshift absorbers can usually be explained as super clustering at high redshift or as gas that is intrinsic to the quasar (Ganguly, Charlton \& Bond 2001b; Richards et al. 2002). For J0027--0944, there is no doubt that the complex absorption lines are caused by the intrinsic gas for a strong reason: the absorption lines show time variability. Because we know that it is unpredictable for an intervening absorption to show variability in such a short time (e.g. Hamann et al. 1995). This is why the time variability can be a powerful tool for identifying quasar intrinsic absorption lines (although quasar intrinsic absorption lines are not always time-variable; e.g. Barlow \& Sargent 1997; Hamann et al. 1997). Especially for the cases in moderate-resolution spectra, which can identify the intrinsic absorption lines neither by partial coverage nor photoionization simulations. 

Furthermore, the coordinated weakening among different absorption components within the \ion{Si}{iv} BAL trough is detected (see Fig. \ref{fig.2} and Table \ref{tab.1}). The different NAL components within the \ion{C}{iv} BAL trough are blended severely, resulting in a P-Cygni absorption though. However, the global weakening does occur across the entire \ion{C}{iv} BAL trough rather than in small segments (see Fig. \ref{fig.1}), which means that the variations of different NAL components within the \ion{C}{iv} BAL are also coordinated. Based on the above, variations of the same absorber of different ions for \ion{C}{iv} and \ion{Si}{iv} are also coordinated. Such well-coordinated variations can be interpreted as a result of global changes in the ionization state of the absorbing gases (e.g. Hamann et al. 2011; Chen \& Qin 2015; Wang et al. 2015).

Line-locking is usually considered as evidence for radiative acceleration (e.g. Foltz et al. 1987; Bowler et al. 2014). Line-locking requires that our sight lines are roughly parallel to the outflow gas motion. The \ion{Si}{iv} BAL in J0027-0944 does not show a line-locking, which means that our line of sight is less likely to be parallel to the outflow wind.

The absorption depths of \ion{Si}{iv} and \ion{C}{iv} BALs are deeper than the corresponding broad emission lines (Fig. \ref{fig.1}), which suggests that the absorbers cover both the continuum source and broad emission line region (BELR). Thus, their distance from the flux source should be larger than the size of the BELR.

\subsection{Two BAL types}% 3.2 %%%%%%%%%%%%%%%
Some of the BALs are believed to have intrinsically diffuse and smooth line profile, they cannot be resolved into multiple discrete narrow components (hereafter Type I BAL, e.g. Hamann et al. 1997; Capellupo et al. 2012). However, in J0027-0944, we did find another type of BAL, consisting of a number of discrete narrow components (hereafter Type II BAL). In other words, the \ion{Si}{iv} and \ion{C}{iv} BALs in J0027-0944 are same as intrinsic NALs in terms of their absorption profiles. On the relationship between the BAL and NAL, there are two main kinds of conjectures: the evolution model and the inclination model (as described in the introduction). Two types of BALs are less likely to evolve into each other (due, for example, to a change of ionization condition, a gas motion, and/or any other mechanisms) because their appearance are completely different. Thus, the evolution model is less likely as the interpretation of such a phenomenon. Compared to the evolution model, the inclination model is favourable `at least' to explain the existence of Type II BAL. In the inclination model, Type I BALs are generally considered to be formed in the main body of the outflow near the plane of accretion disc, while NALs are formed along the line of sight that skim the edges of the BAL flow at higher latitudes above the disc (e.g. Ganguly et al. 2001a; Hamann et al. 2012). While Type II BAL may form in the transitional zone of outflows between Type I BAL and NAL. If the above conjecture is true, then Type II BALs may have the same origin as mini-BALs (see fig. 4 of Hamann et al. 2012).

\subsection{Profile shapes}% 3.3 %%%%%%%%%%%%%%%

As described in the introduction, BALs exhibit a wide variety of profile shapes. In J0027-0944, although having the same origin from the same clumpy structures, the appearance of the \ion{Si}{iv} BAL is different  from \ion{C}{iv} BAL. First, the \ion{Si}{iv} BAL covers a wider velocity range but has weaker EW than \ion{C}{iv} BAL (see Table \ref{tab.1}). Secondly, the \ion{Si}{iv} BAL shows a shape of multiple absorption troughs, while the \ion{C}{iv} BAL shows P-Cygni (see Fig. \ref{fig.1}). These differences may be due to the following reasons. On one hand, due to the difference in fine structure, the red and blue lines of \ion{Si}{iv} $\lambda\lambda$ 1393, 1402 doublets separate farther than those of \ion{C}{iv} $\lambda\lambda$ 1548, 1551 doublets. On the other hand, due to the difference in oscillator strength, the EW/column density of \ion{Si}{iv} NALs tends to be weaker than \ion{C}{iv} NALs. Thus, although they probably have the same origin, the \ion{Si}{iv} BAL still shows multiple absorption features in a wider velocity range, while the \ion{C}{iv} BAL is more severely blended to be a classic P-Cygni in a narrower velocity range. Therefore, the different shapes between the \ion{Si}{iv} and \ion{C}{iv} BALs in J0027-0944 can simply be explained by how crowded the NALs are in velocity space, which depends on the fundamental parameters from atomic physics between \ion{Si}{iv} and \ion{C}{iv}. In addition, it is not a special case that the \ion{Si}{iv} and \ion{C}{iv} BALs in the same quasar spectrum show different  appearances, just like that shown in J0027-0944. 
Previous study also showed that the \ion{Si}{iv} BALs are generally weaker than \ion{C}{iv} BALs (e.g. Capellupo et al. 2012; Filiz Ak et al. 2013), and usually have a  wider velocity range than \ion{C}{iv} BALs (Capellupo et al. 2012). 
Although both \ion{C}{iv} and \ion{Si}{iv} BALs show a wide variety of profile shapes, the phenomenon shown in J0027-0944 can be easily found in other BAL quasar spectra. We found this phenomenon based on a visual check on a large spectral sample of BAL quasars, during a programme to study the BAL variation (Lu et al. 2017). The proportion of Type II BAL will be given quantitatively in future work.

\subsection{Clumpy structure}% 3.4 %%%%%%%%%%%%%%%
The complex of NALs within the BAL troughs and their coordinated variations in J0027--0944 motivate the idea that these absorptions may arise from clumpy gas clouds with similar locations, kinematics and physical conditions. In fact, clumpy structures of outflows have been proved by many works in different aspects. For example, five high-velocity outflow NALs identified in SDSS J212329.46-005052.9 require five distinct clumpy structures of the outflow with similar physical conditions, characteristic sizes and kinematics (Hamann et al. 2011). To solve the so-called `overionization problem' in quasar and active galactic nucleus  outflows, Hamann et al. (2013) suggested that mini-BAL absorbers may consist of a number of small-scale ($d_{\rm cloud}\la10^{-3}-10^{-4}~\rm pc$) but large-density clouds ($n_{\rm e}\ga10^6-10^7~\rm cm^{-3}$). Joshi et al. (2014) also involved a similar picture to explain the strength and velocity variations of a \ion{C}{iv} BAL in quasar SDSS J085551+375752 and J091127+055054.

How the clumpy structures survive in the quasar outflows is still under debate. The classical outflow model (e.g. Murray et al. 1995; Murray \& Chiang 1997) interpreted that the survival of clumpy structures is due to a shielding medium that is located at the bottom of the outflow. Because the shielding medium blocks most of the quasar's far-UV and X-ray radiations, the clumpy structures can avoid overionization at a much lower gas density. This is supported by the observations that BAL quasars are usually relatively X-ray weak compared to non-BAL quasars (e.g. Green et al. 1995; Brandt et al. 2000). However, other observations of NAL or mini-BAL quasars showed less X-ray absorption (Misawa et al. 2008; Chartas et al. 2009; Gibson et al. 2009b), though radiatively accelerated NAL or mini-BAL outflows also have high speeds and ionizations that are similar to BAL absorbers (e.g. Hamann et al. 2011). Hamann et al. (2011) argued that high gas densities in small outflow substructures allow the clumpy structures to survive without significant radiative shielding. In quasar J0027-0944, the existence of \ion{Si}{iv} and \ion{C}{iv} absorption means that these clumpy structures avoid over-ionization, which implies (i) the existence of a shielding medium according to the classical outflow model or (ii) absorbers are self-shielded. In order to further confirm this, it would be interesting to check whether there is a strong X-ray absorption in the quasar J0027-0944. Further study of the X-ray properties of Type II BALs may offer a new insight into the survival mechanism of clumpy structures.

An effective way to solve the clumpy structure of outflows is to use the multiple sightlines caused by gravitationally lensed quasars (e.g. Misawa et al. 2014, 2016). As shown in J0027-0944, for Type II BAL, the decomposition of a BAL into NALs can serve as another way to resolve the clumpy structure of outflows along the line of sight. It is important to resolve a Type II BAL into NAL components. If the high-resolution spectra are obtained, we can measure the covering factors and column densities of this clumpy structure more accurately, which cannot be done via a whole BAL trough. Based on these physical quantities and using the photoionization model, we can further deduce the absorbing region size, the radial distance from the supermassive black hole, the outflow kinetic energy and feedback efficiency.

\section{Conclusion}%% 4 %%%%%%%%%%%%%%%%%%%%%%%%%%%%%%%%%%%
We have found a \ion{C}{iv} BAL that consists of multiple NAL components in the quasar SDSS J0027-0944. Our main results are as the follows.

\begin{enumerate}
 \item Each of the \ion{Si}{iv} and \ion{C}{iv} BALs actually consists of at least four blended NAL systems. In the rest-frame time-scale of about 2.4 yr, all these NAL systems show coordinated time variations (weakening), suggesting that they may originate from the same outflow clouds,  and that their variations can be interpreted as a result of global changes in the ionization state of the absorbing gases.
 \item BALs that consist of a number of NAL components indicate that they are analogous to intrinsic NALs. The existence of two types of BALs prefers the inclination model to the evolution model. This type of BAL, as well as mini-BALs, may be formed at a position between the NALs and diffusion-profile BALs. 
\item The \ion{Si}{iv} and \ion{C}{iv} BALs in J0027-0944 have the same origin but show different profile shapes. The \ion{Si}{iv} BAL shows multiple absorption troughs in a wider velocity range, while the \ion{C}{iv} BAL is P-Cygni in a narrower velocity range. These differences could be interpreted as the substantial differences in their fundamental parameters from atomic physics or just their physical conditions.
\item NAL complex, as one form of BAL, indicates the clumpy structure of this type of BAL outflow. Our discovery offers another way to resolve the clumpy structure for the outflow, which is useful for learning about the characteristics of the clumpy structure of outflows, and for testing the outflow model, when combined to use the high-resolution spectra of the quasar and photoionization model.
\end{enumerate}

This paper presents a discovery that some of the BALs actually consist of blended NALs, but what is the proportion of this type of BALs is not clear. In the future work, we will process a systematic study on the BALs that consist of blended NALs, with the large spectroscopic data set of SDSS. In addition, for some typical cases, we will utilize high-resolution spectra and  photoionization models to study the outflow in more detail.

\section*{Acknowledgements}
We gratefully thank the anonymous referee for many comments that greatly improved the quality of this article. This work was supported by the National Natural Science Foundation of China (No. 11363001; No. 11763001), and the Guangxi Natural Science Foundation (2015GXNSFBA139004).

Funding for SDSS-III was provided by the Alfred P. Sloan Foundation, the
Participating Institutions, the National Science Foundation, and the US
Department of Energy Office of Science. The SDSS-III web site is
\url{http://www.sdss3.org/.}

SDSS-III is managed by the Astrophysical Research Consortium for the
Participating Institutions of the SDSS-III Collaboration, including the
University of Arizona, the Brazilian Participation Group, Brookhaven National Laboratory, Carnegie Mellon University, University of Florida, the French Participation Group, the German Participation Group, Harvard University, the Instituto de Astrofisica de Canarias, the Michigan State/Notre Dame/JINA Participation Group, Johns Hopkins University, Lawrence Berkeley National Laboratory, Max Planck Institute for Astrophysics, Max Planck Institute for Extraterrestrial Physics, New Mexico State University, New York University, Ohio State University, Pennsylvania State University, University of Portsmouth, Princeton University, the Spanish Participation Group, University of Tokyo, University of Utah, Vanderbilt University, University of Virginia, University of
Washington, and Yale University.

%%%%%%%%%%%%%%%%%%%%%%%%%%%%%%%%%%%%%%%%%%%%%%%%%%



%\bibliographystyle{mnras}


%%%%%%%%%%%%%%%%%%%%%%%%%%%%%%%%%%%%%%%%%%%%%%%%%%

%%%%%%%%%%%%%%%%% APPENDICES %%%%%%%%%%%%%%%%%%%%%


%%%%%%%%%%%%%%%%%%%%%%%%%%%%%%%%%%%%%%%%%%%%%%%%%%

% Don't change these lines
\bsp    % typesetting comment
\label{lastpage}
\end{document}